\documentstyle[aps,epsf,floats,twocolumn,prb]{revtex}
\def\RE{{\rm Re}}{\bf }
\def\IM{{\rm Im}}
\newcommand{\bfgr}[1]{{\boldmath{\mbox{$#1$}}}}

\begin{document}
\twocolumn[\hsize\textwidth\columnwidth\hsize\csname@twocolumnfalse%
\endcsname

\title{Griffiths effects and quantum critical points in dirty 
superconductors without spin-rotation invariance: One-dimensional 
examples}
\author{Olexei~Motrunich,$^{1}$ Kedar~Damle,$^{1,2}$ and 
David~A.~Huse$^{1}$}
\address{
$^1$ Physics Department, Princeton University, Princeton, NJ 08544\\
$^2$ Physics Department, Harvard University, Cambridge, MA 02138
}
\date{November 10, 2000}
\maketitle

\begin{abstract}
{We introduce a strong-disorder renormalization group (RG) 
approach suitable for investigating the quasiparticle excitations 
of disordered superconductors in which the quasiparticle spin is 
not conserved.  We analyze one-dimensional models with this RG 
and with elementary transfer matrix methods.  We find that
such models with 
broken spin rotation invariance {\it generically} lie in one of 
two topologically distinct localized phases.  Close enough to 
the critical point separating the two phases, the system has 
a power-law divergent low-energy density of states (with a 
non-universal continuously varying power-law) in either phase,
due to quantum Griffiths singularities.  This critical point 
belongs to the same infinite-disorder universality class as 
the one dimensional particle-hole symmetric Anderson localization 
problem, while the Griffiths phases in the vicinity of the 
transition are controlled by lines of strong (but not infinite) 
disorder fixed points terminating in the critical point.}
\end{abstract}

\vskip 0.3 truein
]

\section{Introduction}
Recently, there has been considerable theoretical activity 
concerning the effect of disorder on the quasiparticle
spectrum in dirty superconductors with different pairing 
symmetries.~\cite{sfwSR,sfnoSR}  The basic philosophy underlying 
most of these developments is to start with a weakly disordered 
problem and investigate the fate (in some RG sense) of disorder at 
large length scales using field theoretic methods.~\cite{plcp}
An analysis of this sort depends most crucially on the possible
symmetries of the quasiparticle Hamiltonian, and it is the 
corresponding universal properties that have received most 
attention thus far.  These include the leading effects of 
disorder on the conductivity and density of states 
in the delocalized regime, as well as the universal scaling 
properties of the localization transition.

Our focus in this article is quite different: 
We ask if Griffiths effects, whereby rare configurations of 
the disorder potential over large regions of space give rise to
{\it non-universal} contributions that dominate some low-energy
property, may be important in dirty superconductors.
In particular, are there situations in which such Griffiths 
effects lead to a singular enhancement in the quasiparticle 
density of states near the Fermi energy?
Unfortunately, this intriguing possibility has not attracted 
much attention in previous work.

Such issues are difficult to address within the weak-disorder
framework mentioned above.  Instead, we introduce a RG approach
that is suitable for situations in which the effective
value of disorder becomes large at low energies.  Our basic 
result is simply stated:  It is indeed possible for such 
Griffiths effects to dominate the low-energy properties, 
at least in one-dimensional
models with either triplet pairing, or strong spin-orbit effects, 
or spin-flip scattering off frozen magnetic impurities.   More 
specifically, we show that such models with broken spin 
rotation (SR) invariance {\it generically} lie in one of two 
distinct localized phases.  These two phases are distinguished 
by a topological property (reflected in the presence or absence 
of zero-energy end-states in a large but finite wire) that makes 
it impossible to go smoothly from one phase to the other.  
At the phase transition separating the two, the effective disorder 
grows without bound when viewed on ever-smaller energy scales, 
and the low-energy density of states behaves as 
$\rho(\epsilon) \sim 1/\epsilon|\ln^3 \!\epsilon|$
(Dyson\cite{dyson} singularity).
The localized phases near this transition, on both sides, 
are {\it Griffiths phases} having a power-law singularity 
(with a non-universal exponent) in the low-energy density 
of states; these Griffiths phases are themselves characterized 
by strong (but not infinite) effective disorder in the limit 
of low energies.
[Note that a recent weak-disorder analysis, Refs.~\onlinecite{gruz}
and \onlinecite{mudry}, of this problem has been carried 
out only at criticality.  The results of
Refs.~\onlinecite{gruz,mudry} for the thermal conductance and 
the density of states are consistent with the predictions of our 
RG approach at such a critical point---however, as we show here, 
the generic behavior of the system is localized rather than 
critical.\cite{ilyafoot}]

The detailed scaling properties of the low-energy strong-disorder
critical point and the nearby Griffiths phases are, of course, 
specific to our one-dimensional examples.  However, our RG approach 
is well suited for studying possible Griffiths effects in two 
or more dimensions as well, and some speculations along these 
lines are briefly discussed towards the end of this article.

\section{Physical picture and motivation}
\label{toy}
Before plunging into the details of our analysis, it is
useful to have a heuristic picture of the basic physics
of our $1d$ problems.  To this end, we consider a simple
toy model for a disordered {\it spinless} (more physically, 
{\it spin-polarized triplet}) superconducting `wire' with a
single transverse mode (channel) active.
We model our system by a lattice Bogoliubov-de Gennes (BdG) 
Hamiltonian
\begin{equation}
\hat{H} = \sum_n ( t c_n^\dagger c_{n+1} + 
\Delta c_n^\dagger c_{n+1}^\dagger + {\rm h.c.} ) +
\sum_n \epsilon_n c_n^\dagger c_n \,.
\end{equation}
In this toy model, the nearest neighbor hopping amplitude $t$ 
and the ``p-wave'' pairing amplitude $\Delta$ are real 
constants, while the impurity potential $\epsilon_n$ can take on 
values $V_1$ and $V_2$ with some probabilities $p$ and $1-p$ 
respectively.  Furthermore, we stipulate that $|V_1|<V_c$ 
and $|V_2|>V_c$, where $V_c \equiv 2|t|$.

The significance of the critical value $V_c$ is readily seen 
by solving the pure problem with fixed $\epsilon_n  = V$, i.e.,
with the chemical potential equal to $-V$.  One easily finds 
that there is a gap in the quasiparticle excitation spectrum 
around the Fermi energy for both $|V|>V_c$ and $|V|<V_c$.
However, at the critical point $|V|=V_c$, the system is gapless.
Thus, there are two different gapped phases for our pure system
(the phase with $|V|>V_c$ simply corresponds, in the absence of
the pairing term, to a situation in which the Fermi level has gone 
below the bottom of the band or above the top of the band;
thus, it is essentially a `band insulator').
For our purposes here, the important distinction between the
two phases has to do with low-energy bound states at the
ends of a long but finite wire of length $L$ with free boundary 
conditions.  In the `gapped superconductor' phase with $|V|<V_c$, 
such a wire has a single quasiparticle state below the gap with 
an excitation energy that is exponentially small in the length $L$; 
the corresponding wavefunction has weight only in the vicinity 
of the two ends of the chain (in the language of 
Ref.~\onlinecite{rg}, in the $L\to \infty$ limit, we thus have 
two zero-energy Majorana fermions, one at each end of the chain) .
In the other gapped phase, $|V|>V_c$, there is no such
low-energy quasiparticle state.~\cite{foottop}

Now, imagine a disorder realization in which the potential has 
value $V_1$ throughout the region between, say, 
sites $0$ and $L$, and value $V_2$ out to a distance 
$L$ on either side of this central segment.  The value
of the potential is left unspecified in the rest of the system.
The probability of this happening is $p^L (1-p)^{2L} = e^{-cL}$,
with $c$ appropriately defined.  Now, the central region can be
thought of, for large $L$, as a finite wire in the phase
with $|V|<V_c$ surrounded by {\it vacuum} (this is reasonable
since the long segments on either side can be roughly thought of 
as regions with no particle because the Fermi energy $-V_2$ has 
gone below the bottom of the band; effective couplings of the 
central region with the rest of the system, mediated through 
such isolating segments, are exponentially small in $L$).
Such a situation will lead to a low-energy quasiparticle
state with excitation energy $\epsilon_L \sim e^{-c' L}$,
with some $c'$ of order one.  Since such low-energy states 
living on `domain walls' between large regions in `opposite' 
phases can happen anywhere along the entire length of our wire,
such disorder configurations will give a non-zero contribution 
in the thermodynamic limit to the density of states.  
This contribution can be estimated as 
$\rho_{\rm Griff}(\epsilon) = \int dL \; 
\delta(\epsilon-e^{-c'L}) e^{-cL} \sim \epsilon^{-1+1/z}$, 
where we have introduced the dynamical exponent $z =c'/c$.
This serves as a lower bound on the actual low-energy density 
of states in our disordered problem---thus, we {\it generically} 
have a power-law behavior (with a non-universal continuously varying
exponent) of the density of low-energy excitations in our toy model.
Although this model is admittedly crude, the picture of rare 
configurations of disorder over large regions of space leading 
to singular low-energy behavior is at the heart of the more 
precise strong-disorder RG analysis of Section~\ref{rsrgnum}.

We conclude this section with some comments on our choice of 
toy-model, and, more generally, on the results obtained in this 
article.  Firstly, note that we completely ignored the 
self-consistency condition that, in principle, determines 
$\Delta$ in terms of the other parameters.~\cite{grt,abr}
This is not expected to matter; in fact, the precise 
choices made for various parameter values are not 
very important for our conclusion---nor is it important
that our toy-model has time reversal symmetry.
As far as this model is concerned, the only important thing is 
the existence of two different gapped phases, with one 
of them supporting zero-energy end-states.
[Note that this is the main distinction of our models without 
SR-invariance from the models with SR-invariance:
When the quasiparticle spin is a good quantum number,
there are no such end-states in any finite open chain,
and we expect no Griffiths effects in this case---this
will also become clear from our more detailed RG analysis.]

Another concern is that we are treating a one-dimensional
superconductor with the BdG equations, which ignore quantum 
fluctuations of the condensate order parameter $\Delta$.  
When the superconductor is in more than one dimension,
$\Delta$ does have a nonvanishing static component at zero 
temperature, but in a strictly one-dimensional system, divergent 
quantum fluctuations mean that the superconducting state does 
not have true long range order or a gap.  However, in highly 
anisotropic quasi-one-dimensional superconductors, $T_c$ 
(and the gap) can, in principle, be large compared to the interchain 
hopping energy,  and our approach should then apply in the range 
between these two energy scales (while no such regime 
appears to exist in quasi-one-dimensional
superconductors known so far, {\em a priori},
there are no physical reasons that would prevent this from 
happening in some cases\cite{schulz}).  Another possible physical 
realization\cite{sfwSR} is that of a vortex in a three dimensional 
gapped superconductor in the presence of frozen magnetic impurities 
or spin-orbit scattering.  In such a situation, the effect of 
disorder on the quasiparticle states confined to the 
vortex core can be analyzed by a $1d$ BdG equation approach such 
as the one we employ.  Naturally, the choice of probability 
distributions for various quenched random variables will
be different depending on the physical realization one is 
interested in (for instance, it is more natural to use a 
quenched random $\Delta$ with zero mean when considering 
the vortex problem).  However, as will be clear from our later 
analysis, the precise form of the probability distribution 
for various bare couplings in the problem does not play an 
important role in determining the nature of the low-energy physics.

Finally, note that the effects of the residual quasiparticle 
interactions are beyond the scope of our analysis.

\section{Formalism, symmetry classes, and our $1d$ models}
It is useful to set up notation and review some 
basics\cite{az,sfnoSR} before proceeding to our actual 
calculations.  To this end, consider a general lattice 
BdG Hamiltonian 
\begin{equation}
\hat{H}_{SC} = \sum_{\alpha \beta} \left(
h_{\alpha \beta} c_\alpha^\dagger c_\beta +
\frac{1}{2} \Delta_{\alpha \beta} c_\alpha^\dagger c_\beta^\dagger +
\frac{1}{2} \Delta_{\alpha \beta}^* c_\beta c_\alpha \right),
\label{HSC}
\end{equation}
where we use a composite label $\alpha=\{i,\mu\}$ for the site
and spin indices of fermion orbitals.  Hopping amplitudes, 
the effects of spin-orbit interaction on the hopping amplitudes, 
spin-flip scattering from frozen magnetic impurities, and random 
potential terms corresponding to non-magnetic impurities are 
now all included in $h_{i\mu, j\nu}$, while pairing amplitudes 
are represented by $\Delta_{i\mu, j\nu}$.
Hermiticity requires $h_{\beta \alpha}=h_{\alpha \beta}^*$, 
and we choose $\Delta_{\beta \alpha}= -\Delta_{\alpha \beta}$
consistent with the fermion anticommutation relations.
Additional restrictions (to be reviewed below) arise when 
T-invariance is a good symmetry (we will {\em not} consider cases 
with SR-invariance in this article).

The spectrum of quasiparticle excitations for $\hat{H}_{SC}$ is 
specified by the  spectrum of a (Hermitian) matrix
\begin{equation}
{\cal H} = \pmatrix{ h         & \Delta \cr
                     -\Delta^* & -h^* }
\label{HBdG}
\end{equation}
acting in an enlarged ``particle-hole'' Hilbert space.
A particle/hole mixing unitary transformation
$U_0= \frac{1}{\sqrt{2}} 
\pmatrix{1 & -i \cr 1 & i} \otimes {\mathbf 1}_{2N}$
(where $N$ is the number of lattice sites), which acts 
independently on states corresponding to each $\alpha$, 
transforms ${\cal H}$ into an antisymmetric 
{\it pure imaginary} form
\begin{equation}
{\cal H}_\IM = U_0^{-1} {\cal H} U_0 =
\pmatrix{ i\, \IM(h + \Delta)  &  i\, \RE(-h + \Delta) \cr
          i\, \RE(h + \Delta)  &  i\, \IM( h - \Delta) }.
\label{HIm}
\end{equation}

This representation is well-suited for a discussion of
heat transport properties of the quasiparticles.  Indeed,
the quasiparticle thermal conductivity is simply 
proportional to $k_B T \sigma$, where $\sigma$ is the $T=0$ 
conductivity of a (normal) system of non-interacting fermions 
described by the lattice Schroedinger equation corresponding to 
${\cal H}_\IM$ (see Ref.~\onlinecite{sfnoSR} and references therein).
[Note that ${\cal H}_\IM$ may also be obtained, as in 
Ref.~\onlinecite{sfnoSR}, from the original $\hat H_{SC}$ by 
writing everything in terms of Majorana fermions and then 
doubling the system, and we will therefore use a ``copy index'' 
$K$ below to label the different blocks of ${\cal H}_\IM$.]

\begin{figure}
\epsfxsize=\columnwidth
\centerline{\epsffile{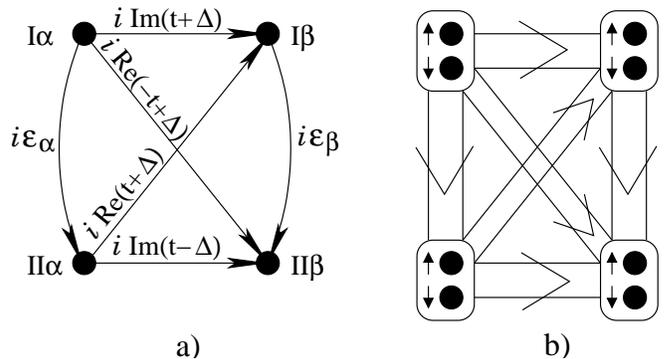}}
\vspace{0.15in}
\caption{ a) General pure imaginary representation of 
hopping ($t$) and pairing ($\Delta$) couplings between
two fermion orbitals $\alpha$ and $\beta$; b) in the
T-invariant case, it is more convenient (see text) to work 
with block-sites and block-couplings shown here. 
}
\label{ImRHfig}
\end{figure}

Most of our discussion will use this pure imaginary form.
In the absence of both SR-invariance and T-invariance,
the different matrix elements of ${\cal H}_\IM$ take on 
roughly independent imaginary values (apart from the 
requirements imposed by hermiticity).  Thus, we have a 
general {\it pure imaginary} random hopping (ImRH) problem on 
a {\it doubled} lattice.  Each original fermion orbital 
$\alpha=\{i,\mu\}$ is ``represented'' by two ``copies'' 
$\mathrm{I}\alpha$ and $\mathrm{II}\alpha$; the energy of 
an orbital $\alpha$ is represented by an imaginary hopping 
amplitude between $\mathrm{I}\alpha$ and $\mathrm{II}\alpha$, 
while hopping and pairing amplitudes between two orbitals 
$\alpha$ and $\beta$ are represented by imaginary hopping 
amplitudes between the two pairs
$\{\mathrm{I}\alpha, \mathrm{II}\alpha\}$ and 
$\{\mathrm{I}\beta, \mathrm{II}\beta\}$ 
(see Fig.~\ref{ImRHfig}a).

\narrowtext

In the presence of T-invariance, $h$ and $\Delta$ satisfy
$\bfgr{\sigma}_y h^* \bfgr{\sigma}_y = h$ and
$\bfgr{\sigma}_y \Delta^* \bfgr{\sigma}_y = \Delta$, where
$\bfgr{\sigma}_y = \sigma_y \otimes {\mathbf 1}_N$ and 
$\sigma_y$ acts on the spin degree of freedom at each site.
The corresponding restrictions on ${\cal H}_{\IM}$ are best
stated as follows:  Group the different spin states on the 
same lattice site $i$ and with the same copy index $K$ into 
a block pair $\{K i \uparrow, K i \downarrow\}$.
T-invariance then implies that there are no internal couplings
within such blocks.  Moreover, the couplings between two blocks 
with the same copy index have a form 
\begin{equation}
\hat t_{\,\mathrm block} = \pmatrix{ i\,a & i\,b \cr i\,b & -i\,a } 
= i r \pmatrix{ \cos\theta & \sin\theta \cr 
                \sin\theta & -\cos\theta }
\label{tTinv}
\end{equation}
with real $a$ and $b$ ($r=\sqrt{a^2+b^2}$), while the couplings 
between two blocks with different copy indices have a form 
$\pmatrix{ i\,c & -i\,d \cr i\,d & i\,c}$ with real $c$ and $d$.
Simply relabeling the spin states of copy~${\mathrm II}$
brings all block-couplings to the same form Eq.~(\ref{tTinv}),
and we will find it convenient to discuss the T-invariant case 
using this picture of block-sites connected by block-couplings
(see Fig.~\ref{ImRHfig}b where each fat arrow represents 
the corresponding block coupling Eq.~(\ref{tTinv}) 
with independent $a$ and $b$, and different arrows are 
roughly independent of each other).  We also use another
representation of ${\cal H}$ in the T-invariant case: 
This is obtained\cite{az} by performing a particle/hole and 
spin-up/down mixing unitary transformation
$U_{\tau} = \frac{1}{\sqrt{2}}({\mathbf 1}_2 \otimes \tau_z +
\sigma_y \otimes \tau_x) \otimes {\mathbf 1}_N$ 
(here, ${\mathbf 1}_2$ and the Pauli matrix $\sigma_y$ act on 
the spin label, while the Pauli matrices $\tau$ act on 
the particle-hole label):
\begin{equation}
{\cal H}_\tau = U_\tau^{-1} {\cal H} U_\tau = 
\pmatrix{0  &  h \bfgr{\sigma}_y - \Delta \cr
         \bfgr{\sigma}_y h + \Delta^*  &  0}.
\label{HTinvBPT}
\end{equation}
${\cal H}_\tau$ thus has a {\it bipartite} complex hopping form, 
which will prove useful for our transfer matrix analysis 
(note however that strong correlations among matrix elements 
of ${\cal H}_\tau$ limit its usefulness in other contexts).

Finally, in the spinless case with no T-invariance, we simply 
drop the spin label altogether and do not have any constraints 
on the bonds of the corresponding ImRH problem.  [Here and
elsewhere in this article, we use the designation `spinless' 
to also refer to all situations in which the spin label of 
the electron can be dropped---such as spin-polarized triplet 
superconductors.  Also, in the rest of this article, we do not 
consider specifically the spinless case with T-invariance, but
only mention the relevant results as we go along.  This case is 
very special~\cite{sfnoSR} and maps onto a class of bipartite hopping problems 
that have been studied both in one\cite{bmf} and two\cite{gade} 
dimensions.  In one dimension, our RG approach yields results
consistent with what is known from Ref.~\onlinecite{bmf} and
can provide more details about the low-energy properties of
any particular system in this universality class, while some 
two-dimensional systems in this universality class will be 
discussed separately.\cite{usgade}]

\vskip 0.2cm
In this paper, we consider three one-dimensional systems:
the spinless superconductor without T-invariance,
the `spinful' T-invariant superconductor, and the `spinful'
superconductor without T-invariance.  In all three examples,
the bulk of our discussion is for the case with a single
transverse mode (channel) present.  However, as will 
become clear from our analysis of the `spinful' superconductor 
without T-invariance, our basic conclusions regarding the 
low-energy physics {\em apply equally well 
for the multi-channel case of all three problems}.

We model these systems by the appropriate BdG Hamiltonians with 
only nearest neighbor hopping and pairing amplitudes, in addition to 
same--lattice-site terms (the restriction to such nearest 
neighbor models is not at all crucial for any of our conclusions).
The ImRH problem corresponding to the 
single-channel spinless case is a
two-leg ladder with all couplings pure-imaginary and roughly
independent, but no other restrictions---in particular, couplings 
along the rungs of the ladder are allowed.  Such `rung-couplings'
(which we will sometimes refer to as `vertical' couplings---see
Fig.~\ref{ImRHfig}a) correspond to any on-site terms in the original 
lattice BdG Hamiltonian; in the spinless case, these can only be 
random potential terms, but more generally, one can also 
have ``s-wave'' pairing amplitudes and spin-flip scattering 
potentials (the latter due to frozen magnetic impurities).
For the single-channel `spinful' T-invariant case,
the corresponding ImRH problem is an analogous two-leg ladder of 
block-sites with pure imaginary block-couplings that are roughly 
independent of one another---in other words, we have a four-leg 
ladder with this special block structure.  Finally, the 
single-channel `spinful' case without T-invariance is a 
pure-imaginary four-leg ladder with no other restrictions.
These ladder problems are related to the bipartite
random hopping (RH) ladder problems~\cite{bmf} in which rung 
couplings are disallowed but the other hopping amplitudes do not 
have to be pure imaginary---as mentioned earlier, these are in 
the same universality class as spinless superconducting wires 
with T-invariance.
[The connection between this special class of models with 
sublattice symmetry and the more general systems that we study 
here is in fact surprisingly close, as will become apparent from 
our transfer matrix and RG analyses.]

Finally, it is worth emphasizing at this point that our focus 
in all these cases is on the low-energy physics: 
The systems mentioned above will all differ from each other 
at higher energy scales, particularly in the initial 
`diffusive' crossover regime that would be present in
any weakly-disordered problem. However, at energies below this
crossover scale, the effective value of disorder becomes large and
all these systems can be described by a unified physical picture.

\section{Strong-randomness RG approach}
\label{strongrgrules}
In order to go beyond the heuristic ideas of Sec.~\ref{toy},
we need a {\it controlled} approach that works in situations 
with strong Griffiths effects (such that $z$, defined by the 
low-energy behavior $\rho(\epsilon) \sim \epsilon^{-1+1/z}$ of 
the density of states, is large).  Such situations are expected 
to correspond to large values of effective disorder in the 
low-energy limit (in fact, a simple scaling argument indicates 
that the width of the distribution of the logarithms of the 
effective couplings is expected to be of order $z$ in the 
low-energy limit).  We are thus led to formulate a strong-disorder 
RG approach to this problem.

Consider the ImRH Hamiltonian 
${\cal H}_{\IM} = \sum_{ij} t_{ij} |i\rangle \langle j|$ with 
$t_{ji}=t_{ij}^*=-t_{ij}$.  The eigenstates of ${\cal H}_{\IM}$
occur in pairs with energies $\pm \epsilon$, and the
strong-randomness RG proceeds by eliminating, at each
step, such a pair of states with energies at the top and
at the bottom of the band:  One finds the largest (in
absolute value) coupling in the system, say $t_{12}$ connecting 
sites $1$ and $2$; this defines the bandwidth of the problem 
$2\Omega = 2\max\{|t_{ij}|\}$.  If the distribution of the 
couplings is broad, the eigenfunctions of the two-site problem 
${\cal H}_{\IM}[1,2]$ will be good approximations to the 
eigenstates with energies $\pm \Omega$, since the couplings 
$t_{1j}$ and $t_{2j}$ of the pair to the rest of the system 
will typically be much smaller.  These couplings can then be 
treated perturbatively, and eliminating the high-energy states
living on the sites $1$ and $2$ gives us the following
effective couplings between the remaining sites:
\begin{equation}
\label{RGrule}
\tilde t_{ij} = t_{ij} - t_{i1} (t_{12}^\dagger)^{-1} t_{2j}
                       - t_{i2} (t_{21}^\dagger)^{-1} t_{1j}.
\end{equation}
Clearly, the renormalized Hamiltonian $\tilde{\cal H}_{\IM}$ 
again corresponds to a pure imaginary hopping problem, but with 
two fewer sites; in particular, the matrix $\tilde{\cal H}_{\IM}$ 
{\it has no diagonal terms},\cite{Rgcomment} 
$\tilde t_{ii} \equiv 0$.

\vspace{0.2cm}
Some remarks on the proposed RG approach are in order here.
Note that our RG rule Eq.~(\ref{RGrule}) is an {\it exact} 
transformation for the zero-energy wavefunction, and as such 
provides information on the zero-energy localization properties.
From a numerical point of view, it can be viewed as a construction 
of the zero-energy wavefunction in an {\it a priori} stable manner.
Moreover, this transformation can also be viewed as an 
approximate but accurate scheme for evaluating the 
`Sturm sequence', i.e., the integrated density 
at very low energies, with the approximations involved being 
well-controlled when the low-energy effective couplings are 
broadly distributed.
Lastly, note that in the limit of strong disorder and in
the absence of strong correlations between different couplings, 
the right hand-side of~Eq.~(\ref{RGrule}) can be replaced with 
the ``max'' (in absolute value) of the three terms.

\vspace{0.2cm}
So far, we have ignored the restrictions that would be imposed 
on the couplings of the ImRH problem in a `spinful' ({\it i.e}, 
when both spin species need to be considered) T-invariant 
situation.  In this case, it is more natural to work with the 
block-sites and block-couplings defined earlier 
(Fig.~\ref{ImRHfig}b).  To begin with, note that the eigenstates 
now occur in {\it doubly degenerate} pairs with energies
$\pm \epsilon$.  Thinking in terms of blocks automatically
incorporates this degeneracy, since the eigenstates of a
two--block-site problem ${\cal H}_{\tau}[{\mathbf 1,2}]$ come
in such doubly degenerate pairs with energies $\pm r_{\mathbf 12} 
= \pm \sqrt{a_{\mathbf 12}^2+b_{\mathbf 12}^2}$.
Our RG approach now eliminates four states at each step, 
two at the top and two at the bottom of the band with energies 
$\pm \Omega$, where $\Omega \!\equiv\! \max\{r_{\bf ij}\}$.
The resulting effective block-couplings among the remaining 
block-sites are again given by~Eq.~(\ref{RGrule}),
but now each $\hat t_{\mathbf ij}$ is a $2 \!\times\! 2$ matrix
of the form~(\ref{tTinv}).  The effective problem is again 
an ImRH problem in the same block-form, and no block-diagonal 
terms are generated.  For the RG rule~Eq.~(\ref{RGrule}),
the flows of bond-energies $r$ and bond-angles $\theta$
do not separate.  However, for strong disorder and in the
absence of strong correlations (i.e., roughly, when 
the ``max'' RG rule applies) the energy variables $r$ flow exactly
as in the ImRH representation of a spinless superconductor
without T-invariance, while the angle variables simply randomize.
Thus, we expect essentially identical results for this `spinful' 
problem with T-invariance and the corresponding spinless problem 
without T-invariance {\it whenever both flow to strong disorder 
sufficiently rapidly}.

\section{Critical points and Griffiths effects in $1d$: 
Transfer matrix analysis}
Before we go on to our more detailed RG analysis of the 
low-energy properties, it is useful to have a picture of 
the phase diagram in each case.  We use an elementary transfer 
matrix analysis to develop such a picture in terms of
the zero-energy localization properties of the system.
The goal here is to show by direct means 
that {\it generically} all our models are localized, and to
demonstrate that there are critical points representing transitions 
between distinct localized phases; of course, these critical points
can only be accessed by fine-tuning the disorder 
distributions.

In general terms, we are looking at the zero-energy localization
properties of $M$-leg ladder systems governed by a
Schroedinger equation
\begin{equation}
\epsilon \vec\psi_n = \hat t_n \vec\psi_{n+1} + 
\hat t_{n-1}^\dagger \vec\psi_{n-1} +
\hat u_n \vec\psi_n,
\end{equation}
where $\hat t_n$ and $\hat u_n$ are $M \!\times\! M$ matrices, and 
the wavefunction $\vec\psi_n$ is an $M$-dimensional vector 
defined at each rung $n$.  We find it convenient to work with 
the following transfer matrix formulation:
\begin{equation}
\pmatrix{\vec\psi_{n+1} \cr \hat t_n^\dagger \vec\psi_n} =
\pmatrix{ \hat t_n^{-1} (\epsilon- \hat u_n) & - \hat t_n^{-1} \cr
          \hat t_n^\dagger & 0}
\pmatrix{\vec\psi_n \cr \hat t_{n-1}^\dagger \vec\psi_{n-1}};
\label{Tdef}
\end{equation}
this defines elementary transfer matrix $\hat T_n$.

\subsubsection{The bi-partite ladder problems:}
\label{bipartite}
To begin with, consider the bipartite problem mentioned above, 
in which one has $M$ coupled chains with no rung couplings, 
$\hat u_n \equiv 0$.  This system was studied by Brouwer 
{\it et.~al.},\cite{bmf} who found that for both real or complex 
hopping amplitudes there are $M+1$ localized phases separated 
by $M$ dimerization driven delocalized critical points; 
each critical point exhibits a strong Dyson singularity in 
the density of states, 
$\rho(\epsilon) \sim 1/\epsilon |\ln^3 \!\epsilon|$.
Here, we rederive by completely elementary means the existence 
of $M$ delocalized critical points, and also characterize 
the $M+1$ distinct localized phases; the ideas introduced in 
the process will generalize naturally to our superconductor 
systems.

Being bipartite, the two sublattices decouple at $\epsilon = 0$:
\begin{equation}
\hat{\rm T}_{n+1} \hat{\rm T}_n = \pmatrix{ 
- \hat t_{n+1}^{-1} \hat t_n^\dagger & 0 \cr
0 & - \hat t_{n+1}^\dagger \hat t_n^{-1}}.
\end{equation}
We are thus led to study the Lyapunov spectrum of the matrix
products $\hat t_{2k}^{-1} \hat t_{2k-1}^\dagger \dots 
          \hat t_2^{-1}    \hat t_1^\dagger$
and $\hat t_{2k}^\dagger \hat t_{2k-1}^{-1} \dots 
     \hat t_2^\dagger    \hat t_1^{-1}$.
At zero dimerization ($\hat t_{2k}$ and $\hat t_{2k+1}$ distributed
identically) the Lyapunov spectra of both products (consisting 
of $M$ distinct Lyapunov exponents in the general case)
are identical and symmetric around zero, and the full Lyapunov
spectrum (of the full transfer matrix product) is doubly 
degenerate.  Thus, for even $M$ the smallest (in absolute value) 
Lyapunov exponent is non-zero and the zero-energy modes 
are localized, while for odd $M$ there is always 
a zero Lyapunov exponent.  
[The actual values of these exponents are not of much 
concern here, only the fact that they are all distinct.]  
Now, consider adding dimerization by simply 
multiplying every odd bond $\hat t_{2k+1}$ by a scale factor 
$e^{\,\delta}$ and every even bond $\hat t_{2k}$ by a factor 
$e^{-\delta}$.  Clearly, the whole Lyapunov spectrum for one reduced 
(sublattice) problem is shifted rigidly by $\delta$ 
(and by the exactly opposite amount for the other sublattice).  
Thus, as we scan $\delta$ from $-\infty$ to $+\infty$, 
there will be exactly $M$ points where two of these exponents 
of the full transfer matrix cross zero.  For the given sublattice, 
if we label each non-critical region by $(k,M-k)$ with $k$ growing 
modes and $M-k$ decaying modes, these critical points represent 
consecutive ``delocalization'' transitions
\mbox{$(k,M-k) \rightarrow (k+1,M-k-1)$}.  Of course, this
can also be restated in terms of the number of zero-energy
states localized at each end in a finite odd-length chain 
with free boundary conditions, and provides a `topological' 
distinction between the different localized phases; 
each transition corresponds to a single zero-energy state becoming
delocalized and migrating from one end of the chain to the other.  

In a bulk system, at any of these critical points, one has two 
zero-energy delocalized {\it Lyapunov modes} (linear combinations 
of which can roughly be interpreted as a `left-moving' and a
`right-moving' slow modes).  Since any description of the low-energy
properties of the critical system in terms of such `slow' modes
has to respect the bipartite nature of the original problem, it is 
natural to expect that such a low-energy effective theory is in the 
same universality class as the bipartite single chain RH 
problem.\cite{mathur}
This provides a clear rationale for the critical low-energy 
density of states to be of the Dyson form.  Similarly, we expect 
that the localized phases in the vicinity of any critical point 
look, at low energies, like the dimerized Griffiths phases 
obtained by introducing a small amount of dimerization in the 
single chain problem---in particular, we expect Griffiths 
singularities in the density of states consistent with the 
results of Ref.~\onlinecite{bmf}.

\subsubsection{Single-channel spinless superconductor without 
T-invariance:}
Returning to the dirty superconductor problems, consider
first the spinless fermion system with no T-invariance 
in the ImRH language.  In this case, the rung coupling is of 
the form $\hat u_n = \mu_n \sigma_y$ with some real $\mu_n$ 
($\sigma_y$ acts on the copy label $K$),
while the hopping term is a general pure imaginary 
$2 \!\times\! 2$ matrix $\hat t_n$.  Because of the identity 
$\sigma_y \hat a^{-1} \sigma_y = \hat a^T / \det \hat a$
valid for any $2 \!\times\! 2$ matrix $\hat a$, the zero-energy
transfer matrices ``decouple'':
\begin{equation}
\hat{\mathrm T}_n = -\pmatrix{ 1 & 0 \cr 0 & \sigma_y }
\pmatrix{\mu_n \hat t_n^{-1} \sigma_y & \hat t_n^{-1} \sigma_y \cr
         \tau_n \hat t_n^{-1} \sigma_y & 0}
\pmatrix{ 1 & 0 \cr 0 & \sigma_y },
\label{Tdecouple}
\end{equation}
where $\tau_n = \det \hat t_n$.  Thus, Lyapunov exponents of the 
product of $\hat{\mathrm T}_n$ are given by a ``superposition'' 
of the Lyapunov exponents of the product of 
$\sqrt{|\tau_n|} \hat t_n^{-1} \sigma_y$ and of the product of
$\pmatrix{
\mu_n / \sqrt{|\tau_n|} & 1/\sqrt{|\tau_n|} \cr
{\mathrm sign\,}(\tau_n) \sqrt{|\tau_n|} & 0}$.
The former product is very similar to the product 
$\Pi_k \hat t_{2k}^{-1} \hat t_{2k-1}^\dagger$ studied earlier
(note, however, the particular ``normalization'' used here);
the Lyapunov spectrum consists of two exponents $\pm \gamma_t$, 
with $\gamma_t$ of order one.  The latter product is 
essentially the transfer matrix product at $\epsilon=0$
for a $1d$ Anderson problem with random energies $\pm \mu_n$ 
(depending on the sign of $\tau_n$) and random hopping 
amplitudes $\sqrt{|\tau_n|}$; the two corresponding exponents 
are $\pm \gamma_\mu$.  The full Lyapunov spectrum thus consists 
of the four exponents $\pm\gamma_t \pm \gamma_\mu$.
As we increase the strength of the rung couplings $\mu_n$ 
(e.g., by increasing the root mean square strength 
$R \equiv R_\mu \equiv \sqrt{\overline{\mu^2}}$ 
with $\overline\mu = 0$ kept fixed) from $0$ to $\infty$, 
$\gamma_\mu$ also increases from $0$ to $\infty$.
Thus, at some critical strength $R=R_c$ of the rung couplings, 
$\gamma_\mu$ will equal $\gamma_t$, and two Lyapunov exponents 
will be zero corresponding to an isolated delocalized critical 
point along the $R$ axis. 

\begin{figure}
\epsfxsize=3in
\epsfxsize=2.4in
\centerline{\epsffile{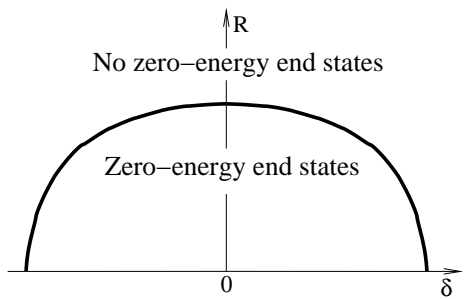}}
\vspace{0.15in}
\caption{Schematic phase diagram in the $\delta-R$ plane
for both the general spinless case without T-invariance,
and the `spinful' model with T-invariance.  We use $R$ 
to denote some measure of the strength of the rung couplings 
which correspond to the random on-site terms of the original 
superconductor problem (e.g., for a symmetric distribution 
of random rung couplings, we can define $R$ to be a root mean 
square of this distribution).  
The vicinity of the phase boundary 
is expected to exhibit strong Griffiths effects. 
}
\label{phased1}
\end{figure}

Now, in the superconductor problem, there are no eigenstates 
at precisely zero energy in any system---this is due to the 
presence of the rung couplings.  Nevertheless, there may be
states with exponentially small (in system size) energy,
and the localized phases on either side of the critical point
can again be characterized in terms of such almost--zero-energy 
end-states.  Consider again an open odd-length chain.
When $\mu_n \equiv 0$ (and $\delta \equiv 0$), there are
two zero-energy states, one at each end of the chain.
Turning on the rung couplings enables the two to mix, but since 
they are separated by the entire length of the chain, the 
splitting is exponentially small in the length of the 
chain---as long as $R < R_c$. 
Thus, in this phase, there will be two such (essentially) 
zero-energy end-states.  In terms of the quasiparticle spectrum 
of the original superconductor $\hat{H}_{SC}$, there is a 
{\it single} quasiparticle state with an exponentially small 
energy and a wavefunction with weight only at the two ends of 
the chain.  On the other hand, in the phase with 
$R > R_c$, there are no such end-states with nearly zero 
energy, as may be argued by 
starting with the limit of large $\mu_n$.
Thus, we have two different localized phases distinguished
by this topological property; if we think in terms of
a more general `dimerization'-`rung coupling'
($\delta-R$) parameter plane, we have a phase diagram 
shown schematically in Fig~\ref{phased1}. 
Moreover, we again expect the low-energy properties in the 
vicinity of the transition between the phases to be in 
the universality class of the single chain bipartite
RH problem in the vicinity of its zero-dimerization critical 
point. In particular, we again expect Griffiths singularities 
in the density of states of either localized phase in the 
vicinity of the critical point---our expectations will be borne 
out by the more detailed RG analysis in the next section.

\narrowtext

\subsubsection{Single-channel spinful T-invariant superconductor:}
\begin{figure}[t]
\epsfxsize=\columnwidth
\centerline{\epsffile{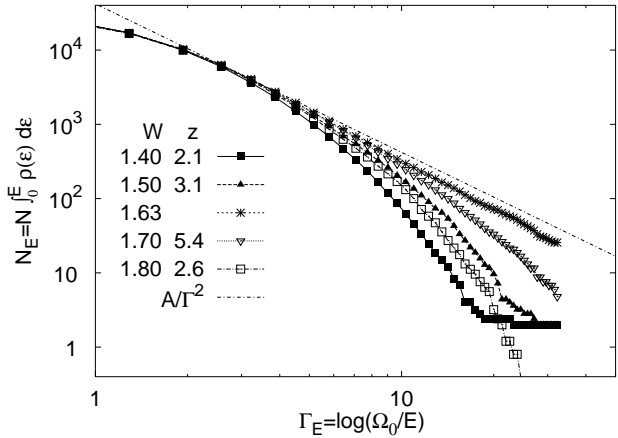}}
\vspace{0.15in}
\caption{Numerical check of the Dyson singularity in the
`spinful' T-invariant case.  Number of states with energies 
between $0$ and $E$ for an open chain of length $L=10^5$, 
averaged over $5$ disorder realizations, is plotted vs 
the log-energy scale $\Gamma_E=\ln(1/E)$.
Independent inter-site couplings (real and imaginary parts, 
of independent hopping and pairing amplitudes) are chosen 
from a uniform distribution over $[-1,1]$, while independent 
on-site couplings (real chemical potential and real singlet 
pairing amplitude)
are chosen from a uniform distribution over 
$[-W,W]$.  Critical $W_c=1.630 \pm 0.001$ was found 
accurately from numerical transfer matrix analysis.
At this point, we clearly have $N_\Gamma \sim 1/\Gamma^2$
(which is shown displaced from the data for clarity).
We also show several off-critical points and give rough
estimates of the corresponding dynamical exponents $z$
from the Griffiths fits $N_E \sim E^{d/z}$ over $4<N_E<1000$.
}
\label{TinvDOS}
\end{figure}

Consider now the `spinful' system with T-invariance (but
no SR-invariance).
In the bipartite representation, Eq.~(\ref{HTinvBPT}), the
rung coupling is
$\hat u_n=\pmatrix{0 & w_n \sigma_y \cr w_n^* \sigma_y & 0}$
with some c-number $w_n$, while the hopping term has a form
$\hat t_n=\pmatrix{0 & \hat a_n \cr \hat b_n & 0}$
with $\hat b_n = -\hat a_n^*$ but otherwise general
complex $2 \!\times\! 2$ matrix $\hat a_n$.  From the
transfer matrix, Eq.~(\ref{Tdef}), the sublattice decoupling
at zero energy is seen immediately; on one sublattice, we
need only consider the product of
\begin{equation}
\hat{\mathrm Q}_n = -
\pmatrix{\hat a_n^{-1} w_n \sigma_y & \hat a_n^{-1} \cr 
         \hat a_n^T & 0}.
\end{equation}
Lyapunov exponents of this product together with their 
negatives (from the other sublattice) form the full Lyapunov 
spectrum of the system.  But this product further ``decouples'' 
precisely as in Eq.~(\ref{Tdecouple}) into the product of
$\sqrt{|\alpha_n|} \hat a_n^{-1} \sigma_y$ and the product of
$\pmatrix{
w_n / \sqrt{|\alpha_n|} & 1/\sqrt{|\alpha_n|} \cr
\alpha_n/ \sqrt{|\alpha_n|} & 0}$; here $\alpha_n=\det \hat a_n$.
Lyapunov exponents of the former product are $\pm \gamma_a$
with some $\gamma_a$ of order one.  However, the $1d$ Anderson 
localization problem corresponding to the latter product is 
non-hermitian: while the hopping amplitudes can be chosen real 
and equal to $\sqrt{|\alpha_n|}$, the on-site energies 
$|w_n| e^{i\psi_n}$ are {\it complex}, with the phases 
$\psi_n$ having contributions from the phases of both
$w$ and $\alpha$.  Nevertheless, the Lyapunov spectrum still
consists of two exponents $\pm \gamma_w$.  The spectrum of 
the product of the $\hat{\mathrm Q}_n$ is thus 
$\pm \gamma_a \pm \gamma_w$, and the other sublattice merely
duplicates this to make the full Lyapunov spectrum doubly 
degenerate.
Now, all we need to know is that for small values of 
rung-couplings the corresponding $\gamma_w$ is also small,
while for large values of rung-couplings $\gamma_w$ is large.
Then, as in the spinless case, there is an isolated delocalized 
critical point along the 
$R \equiv R_w \equiv \sqrt{\overline{w^2}}$ axis 
for some critical strength $R_c$ of the rung coupling terms
(at which $\gamma_w = \gamma_a$).  Note that at this critical 
point, a total of four Lyapunov exponents will simultaneously
cross zero; the corresponding two pairs of critical modes 
are related to each other by $T$-invariance.
Again, the phase with $R < R_c$ is
characterized by the presence of end-states with exponentially 
small energies.  The only difference from the spinless case is 
that there are now four of them---this corresponds, in terms of 
$\hat{H}_{SC}$, to two T-symmetry related quasiparticle states, 
each with an exponentially small energy and a wavefunction having 
weight only at the two ends of the chain.  The phase with
$R > R_c$ again has no such nearly--zero-energy end-states.
In the full $\delta$-$R$ plane, we thus have the schematic phase diagram
shown in Fig~\ref{phased1}.
Again, the critical point and the phases in its vicinity are
expected to look, at low energies, like those in the
single chain RH problem, with an additional degeneracy 
introduced by T-invariance.  A direct numerical check of 
the Dyson form for the critical density of states in this
`spinful' T-invariant case is shown in Fig.~\ref{TinvDOS}, 
while the RG results of the next section confirm the underlying 
physical picture in detail.

\narrowtext

Parenthetically, we also note that these Lyapunov exponent 
crossing arguments imply that in all cases considered so far, the 
inverse of the smallest Lyapunov exponent diverges as 
$|g-g_c|^{-1}$ where $g$ is some tuning parameter that drives 
the system through the critical point at $g_c$.  This implies
that the exponent $\nu$ for the typical localization
length is $\nu=1$ at all these transitions.

\subsubsection{Single-channel spinful superconductor without 
T-invariance:}

Finally, we turn to the `spinful' case with neither T-invariance 
nor SR-invariance (again, with only a single transverse channel).
[Note that a quasi-$1d$ spinless fermion superconductor with
{\it two} transverse channels would also be modeled by the same
BdG problem, with a somewhat different interpretation for the
various couplings.  Thus, analysis of this case is of value
in demonstrating that our basic conclusions are not
special to the single-channel case in any of the problems we 
consider.]

In this case, we have been unable to come up with any simple 
decoupling scheme that maps the corresponding transfer matrix 
to that of some problem with sublattice symmetry---the exact 
mapping we have used earlier is thus special to the two 
single-channel cases considered above.  However, when the rung 
couplings are all zero, we do know that the corresponding bipartite 
four-leg ladder has five phases as we scan the dimerization 
parameter $\delta$---these are labeled $(k,4-k)$ with 
$k=0,1,2,3,4$ corresponding to $4-k$ zero-energy
states localized at one end and $k$ states localized at 
the other end for a finite ladder with an odd number of 
rungs and free boundary conditions.
In particular, in the vicinity of $\delta=0$ one has
two such states at each end.  Now, turning on some weak rung 
couplings allows these two states (at each end) to mix 
amongst themselves, and there will thus be no states with 
exponentially small energies in this regime.  On the other 
hand, the same is clearly true for very large values of 
the rung couplings.  Thus, the phases obtained in either case in
the vicinity of $\delta=0$ are expected to `look' the same (this is made more
precise later using our RG approach).

\begin{figure}
\epsfxsize=\columnwidth
\centerline{\epsffile{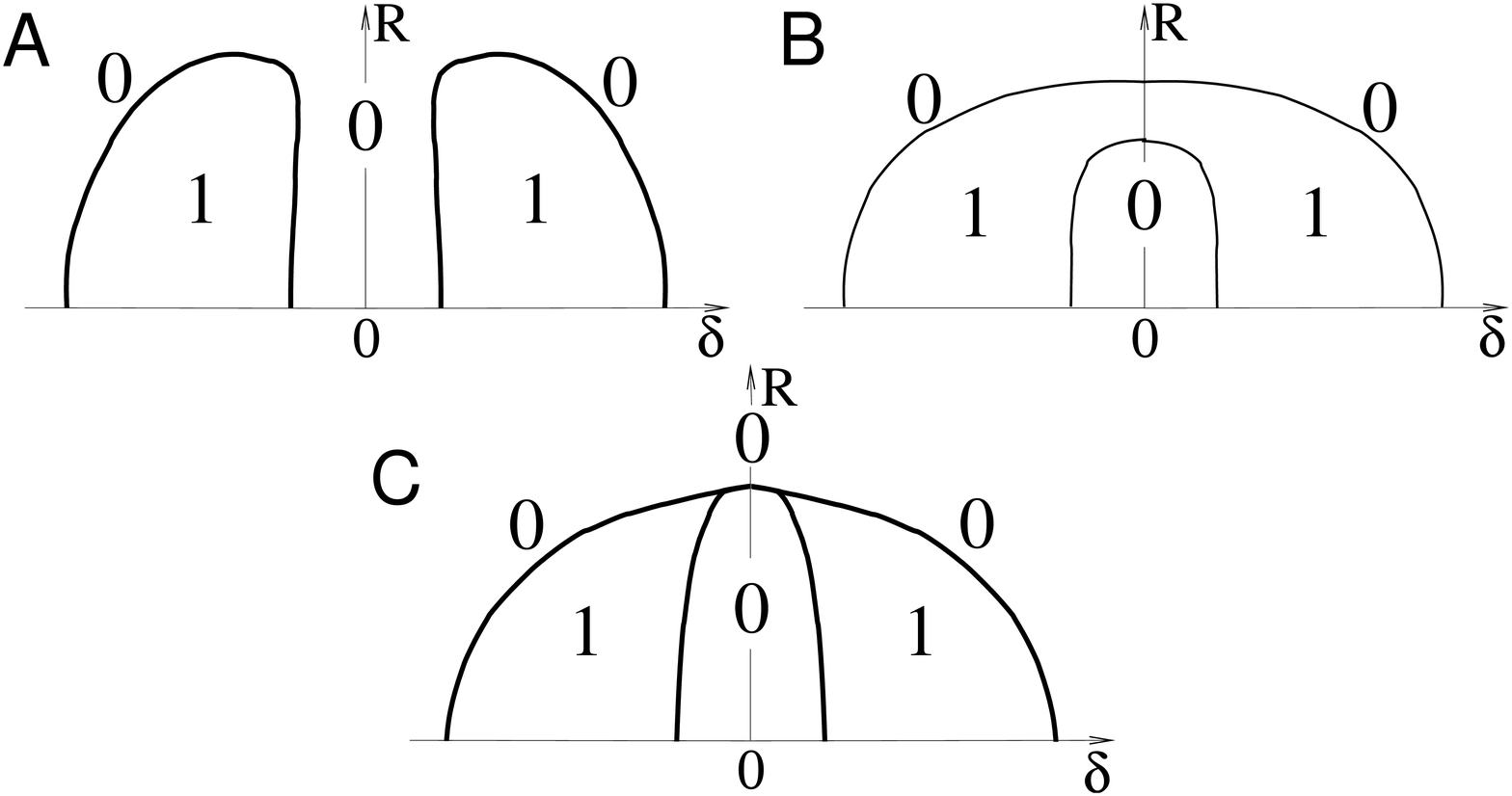}}
\vspace{0.15in}
\caption{Possible phase diagrams in the $\delta-R$ plane
for the `spinful' case without T-symmetry.  We use 
$R$ to represent some measure of the strength of the 
rung-couplings.  On the diagrams, we denote the phases 
with zero-energy end-states by {\tt 1} and the phases with 
no such states by {\tt 0}.  {\tt A} and {\tt B} represent 
two simple possibilities, while a more complicated case with 
a multi-critical point is shown in {\tt C}.  Our numerical 
transfer matrix studies tentatively suggest that in the most 
`random' such superconductor (i.e., with all couplings present 
and independent) the phase diagram is very nearly that of 
the panel {\tt C}.
In many other cases (e.g., when we allow only on-site 
chemical potentials in the original superconductor 
problem, but completely general inter-site couplings), 
we observe the phase diagram {\tt A}.  We have
not found realizations that would clearly exhibit the phase 
diagram {\tt B}; however, we do not have arguments that would 
rule out this possibility.
}
\label{phased2}
\end{figure}

The question then arises:  Is there a single intervening phase 
(at $\delta=0$) which is topologically different? 
This would result in {\it two} phase transitions as we scan 
the magnitude of the rung couplings.  Another simple possibility 
is that there is no transition at all as a function of increasing 
rung coupling.  Of course, there can also be other more 
complicated scenarios (some possibilities for the full phase 
diagram are sketched in Fig.~\ref{phased2}).
Moreover, since there are several independent rung couplings
corresponding to each physical lattice site, there are many 
different ways of `increasing the rung coupling' and the possible
phases and transitions encountered along the way will most likely 
depend on how we scan.

\narrowtext

To get a more detailed picture, we performed an extensive 
numerical transfer matrix analysis (to obtain accurate results, 
we used the technique of Ref.~\onlinecite{chalker}), as well as 
exact diagonalization studies, for particular choices of scan.
In one choice of scan, we include all possible inter-rung 
couplings, but allow only those (intra-)rung couplings that 
correspond to random on-site chemical potentials in the 
original superconductor problem.  In this case, we can
clearly delineate the phase boundaries to conclude that
we have a phase diagram of the type shown in 
Fig.~\ref{phased2}{\tt A}, with no transition along 
the $R$ axis at $\delta=0$ (with $R$ now a measure of the 
strength of the on-site potentials).
If we scan across the phase boundary at a fixed non-zero value 
of dimerization (so that we start in the phase with zero-energy
end-states and leave it by increasing the mean-square strength 
of the random potentials), we find that the typical localization 
length defined by the inverse Lyapunov exponent again diverges 
with an exponent $\nu=1$, and the critical density of states
is again of the Dyson form.

In another scan, we take the most random such system (i.e., with 
all allowed couplings present and independent) and boost the 
strength of all the (independent) rung couplings by the same 
amount on average, while keeping distribution of the inter-rung 
couplings fixed.  
As we increase the typical magnitudes of the rung couplings 
from zero in this manner, we again see strong Griffiths 
singularities in the density of states developing, and, 
possibly, critical behavior.  However, near such a tentative 
critical point the two Lyapunov exponents that come close 
to zero seem to have a mutual ``level repulsion'' and 
the Lyapunov spectrum seems to exhibit the analog of an 
``avoided level crossing''; this results in huge localization 
lengths and strong near-critical behavior in large regions 
around the tentative critical point.
A more detailed investigation away from $\delta=0$ yields
a phase diagram of the type shown in Fig.~\ref{phased2}{\tt C}.
Thus, to within our numerical accuracy, there seems 
to be a multicritical point in the $\delta-R$ phase diagram at
$\delta=0$.  However, we can not exclude the possibility that 
we are seeing a case with no transition along the $R$-axis 
(but phase-boundaries coming very close to this axis) or an 
almost-degenerate case with two very closely spaced transitions.  
Crossing the phase boundaries away from the putative 
multi-critical point again gives a localization length exponent 
of $\nu = 1$.  However, we are unable to make any reliable statements
in the vicinity of the multicritical point.

We have scanned along several other directions in the
parameter space (corresponding to different interpretations 
of the rung-coupling parameter $R$), but have not clearly
seen two distinct transitions as in Fig.~\ref{phased2}{\tt B}; 
however, behavior of the type shown in Fig.~\ref{phased2}{\tt A} 
is the most common.[However, note that the weak-disorder analysis of 
Ref.~\onlinecite{gruz} did find critical behavior in 
the conductance at $\delta=0$ with inter-rung 
and intra-rung couplings identically distributed and chosen
from a Gaussian distribution. Moreover, their
result is consistent with our predictions for critical behaviour
at `ordinary' critical points, as opposed to multicritical points.]

These numerical results thus confirm our suspicion that 
the structure of the full phase diagram in this general 
`spinful' case (or in multi-channel versions of all the cases) 
is quite complicated.  The specific phase diagram obtained 
by tuning the parameters of a particular physical system 
(in which some subset of the allowed couplings may be 
identically zero) can thus be very different from case 
to case (a similar observation in related $2d$ models has 
been made in Refs.~\onlinecite{chalker}~and~\onlinecite{rl}).
However, it is clear that there can be, in general, two kinds
of localized phases.  Moreover, whenever both phases
are present in the phase diagram of a particular system, 
we again expect
(analogous to the single-channel cases considered above)
the system in the vicinity of the phase boundary to `look'
at low-energies like a single-chain random hopping problem with
weak dimerization (note that the numerical estimate, $\nu = 1$,
that we obtain away from any special `multicritical' points 
supports this picture, and our RG results provide further 
confirmation of the same).

This is as far as we can go with an elementary transfer matrix
analysis.  For a more detailed characterization of the low-energy
properties, we now turn to the strong disorder RG analysis.

\section{Critical points and Griffiths effects in $1d$: RG analysis}
\label{rsrgnum}
To test the above picture in detail, we have implemented 
the RG numerically in all three single-channel cases. 
Since the single-channel spinful problem without T-invariance is 
essentially identical to the corresponding two-channel spinless 
problem, this last example also serves to establish that our 
conclusions do not depend on any special properties of the 
single-channel case.  In the single-channel spinless case without 
T-symmetry (and the corresponding spinful case with T-invariance), 
we focus mainly on the vicinity of the phase transition 
at $\delta=0$ (see Fig.~\ref{phased1}).  In the `spinful' case 
without T-invariance, we consider the two different realizations 
described in the
previous section---these have phase diagrams of the types shown 
in Fig.~\ref{phased2}{\tt A} (in which we scan across the phase
boundary at non-zero dimerization so that we leave the phase 
having end-states by increasing the strength of the onsite 
potentials) and in Fig.~\ref{phased2}{\tt C} (in this case, we 
focus on the immediate vicinity of the putative multicritical 
point).  [For completeness, we have also studied the transitions 
as a function of dimerization in the bipartite ladder problems
of Sec.~\ref{bipartite}.]

For the $\delta=0$ spinless case, (and the `spinful' T-invariant
case) we consider systems with lengths as large as $L=2 \times 10^5$.
The initial conditions used have random (inter-rung) hoppings 
chosen from a uniform distribution over $[-1,1]$, and
random (intra-)rung couplings $u$ chosen from a symmetric 
(with either sign equally probable) power law distribution
$P(u)=\frac{1}{2g} |u|^{-1+1/g}$, $|u| \leq 1$.
Note that the RG transformation~(\ref{RGrule}) can be 
formulated entirely in terms of the imaginary parts of 
the couplings, and this is the language that we use here.
For the `spinful' case without T-symmetry, we are restricted
to $L \leq 5 \times 10^4$.  When we scan at finite dimerization,
we enforce this dimerization by choosing the even and odd 
inter-rung couplings from the power law distribution $P(u)$, but
with different fixed values $g_{\rm even} \neq g_{\rm odd}$ for 
the power law exponents.  The strength of the rung couplings is 
again tuned by varying the corresponding power-law exponent 
$g_{\rm rung}$.
We use the ``full'' RG rule~(\ref{RGrule}) rather than its less 
accurate ``max'' version since we are primarily interested in 
testing our physical picture for the low-energy properties 
starting with a system with moderate values of the bare disorder.

Apart from the immediate vicinity of the putative multicritical
point of Fig.~\ref{phased2}{\tt C} (which we comment upon 
separately), our results are equally reliable and essentially 
identical in all the cases studied.  In the interests of brevity, 
below we display in detail only the results obtained in the spinless 
case at $\delta=0$ (see~Fig.~\ref{phased1}).

\narrowtext
\begin{figure}
\epsfxsize=\columnwidth
\centerline{\epsffile{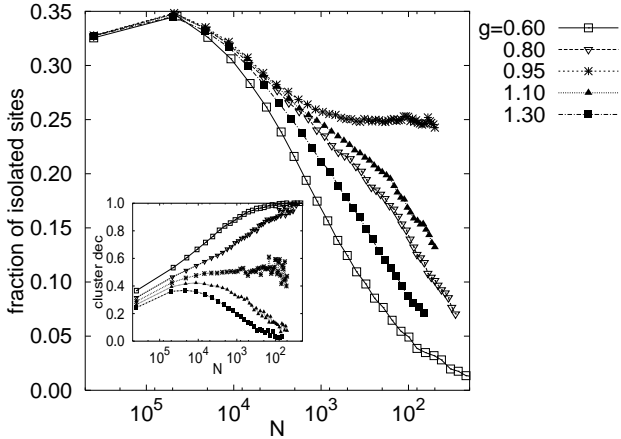}}
\vspace{0.15in}
\caption{Fraction of isolated sites among the remaining
sites for the two-leg ImRH ladder representing spinless fermion
superconducting wire.  Critical $g_c \approx 0.95$ is
clearly identified.  Inset shows the fraction of cluster 
decimations (see text for details).
}
\label{RGKM2isolani}
\end{figure}

We search for the critical point by looking at the fraction 
of ``isolated'' sites among the remaining sites in the system:  
If there are $N$ sites left, we find all the sites that are 
``covered'' by the $N/2$ strongest remaining bonds.  
Roughly speaking, the fate of these covered sites is 
clear---they will be frozen dynamically at rougly this log-energy 
scale.  The rest of the sites are still dynamically free at 
this scale, and we call them ``isolated''.  In a localized 
phase, this fraction quickly approaches zero.  On the other
hand, at a critical point, we expect that new couplings which
contribute, upon their subsequent decimation at a lower energy, to the density
of states at that lower scale, are formed continually over all
energy scales in a scale-invariant way.
The fraction of isolated sites at criticality is therefore expected
to saturate to some constant at low energies.
In the case of a single critical RH chain, a quarter of the 
remaining sites is notionally isolated at each stage of the RG: 
each site has a bond to the right and to the left, and the bond 
strengths are uncorrelated; since each bond has a $50\%$ chance of 
being ``weak'', the site is ``isolated'' with probability $1/4$.  
This gives us a very direct test for the location and
nature of the critical point.

Results of such a search for our spinless case are shown on 
Fig.~\ref{RGKM2isolani}.
We clearly identify a critical $g_c \approx 0.95$, and note
that the fraction of isolated sites at this critical point
remains essentially $0.25$, which is evidence that the low-energy
behavior is that of a single one-dimensional `backbone'
that goes critical.  Note that it is also possible to further probe 
the geometry associated with the low-energy theory, as is done, 
e.g., below in our ``order-parameter'' studies, and even more 
exhaustively by looking in detail at probability distributions 
of the various couplings (which we have not pursued fully---the RG 
results shown here already confirm our basic picture of the 
low-energy physics).

\begin{figure}[t]
\epsfxsize=\columnwidth
\centerline{\epsffile{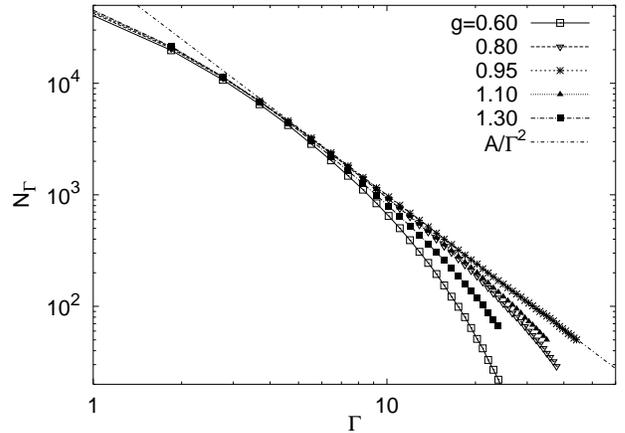}}
\vspace{0.15in}
\caption{Number of remaining sites $N$ vs log-energy scale
$\Gamma$.  Also shown is the fit of the critical $N_\Gamma$
to the form $A/\Gamma^2$.
}
\label{RGKM2NvsG}
\end{figure}

\begin{figure}[t]
\epsfxsize=\columnwidth
\centerline{\epsffile{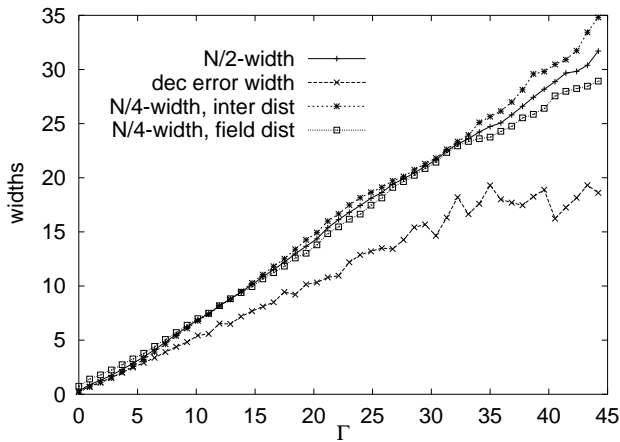}}
\vspace{0.15in}
\caption{At the critical point (determined from 
Fig.~\ref{RGKM2isolani}) different measures of the
widths of the {\it log-}coupling distributions all scale
linearly with $\Gamma$.  We plotted the $N/2$-width of
the distribution of all bonds, the `decimation error'
width (defined as the average logarithm of the ratio of
the decimated bond to the strongest nearby bond), and
also the $N/4$-widths of the interaction and field 
distributions with respect to the vertical reference
cover (see text for details).
}
\label{RGKM2CritW}
\end{figure}

At the critical point, the number of sites remaining at the
log-energy scale $\Gamma \equiv \ln(\Omega_0/\Omega)$ is 
$N_\Gamma \sim 1/\Gamma^2$, as can be readily seen from 
Fig.~\ref{RGKM2NvsG}.  Since $N_\Gamma$ is essentially 
the integrated density of states, this is consistent with
the Dyson $\rho(\epsilon) \sim 1/\epsilon |\ln^3 \!\epsilon|$.
Moreover, at $g=g_c$, the distributions of all couplings 
become broader and broader on the {\it logarithmic} scale, 
with the characteristic widths growing linearly with $\Gamma$.  
This is shown on Fig.~\ref{RGKM2CritW}.
In conjunction with our result for the fraction of isolated
sites at low energies at criticality, this indicates
that the critical point is in the same universality class
as the single chain RH problem.  [Note that since the system 
effectively reduces to a single RH chain `back-bone' in the 
low-energy limit, we expect\cite{mathur} the critical average 
thermal conductance $\kappa_c(L)$ of the original superconductor 
to scale as $k_BT/\sqrt{L}$, where $L$ is the length of the 
wire---this is consistent with the weak disorder result of 
Ref.~\onlinecite{gruz}.  As mentioned earlier, this leads us 
to believe that their analysis was performed only at 
such a critical point and does not represent the generic 
behavior of the system.]

\narrowtext

\narrowtext

Consider now the two different localized Griffiths phases.  
As expected, we find $N_\Gamma \sim e^{-\Gamma/z}$ with a 
continuously varying dynamical exponent $z$ which diverges 
as $z \sim |g-g_c|^{-1}$ (to within our numerical accuracy)
as we tune across the critical point.  Within our strong-disorder approach, 
the two Griffiths 
phases are distinguished by the character of the corresponding 
RG-generated dimer patterns.  To discuss this intuitive
distinction more precisely, we compare such RG-generated
dimer covers against some fixed {\it reference dimer cover}.
A natural choice of the reference cover for the spinless 
problem at hand is a ``vertical'' cover with reference dimers 
covering the rungs of the ladder, i.e., joining the 
two copies of each of the original fermions of $\hat H_{SC}$.
We use this specific reference cover in our numerical studies 
presented below; the following discussion, however, is fairly 
general.  [We use the same `vertical' reference cover in the 
other two superconductor problems we study with the RG.]

We call two sites connected by such a reference bond
a {\it cluster}, and the corresponding coupling between the
two sites a {\it field} on the cluster.  Couplings that
connect sites in different clusters are called 
{\it interactions} between the clusters.  This terminology
is borrowed from the $1d$ random transverse field Ising model 
(RTFIM), but we emphasize that the correspondence is {\it not} 
exact although we do expect that the {\it critical} behavior,
characterized with respect to a well-chosen cover, is 
essentially that of the RTFIM (we expect this to be true 
because a similar analysis for the dimerized single chain
random hopping problem using the natural reference cover 
consisting of alternate bonds gives an exact mapping to
the $1d$ RTFIM).  When a coupling connecting 
two sites in the same cluster is decimated (field decimation), 
the corresponding cluster is {\it killed}.  When a coupling 
connecting two sites in different clusters is decimated 
(interaction decimation), the two clusters are {\it joined} 
into one new cluster, which is now specified by the two other 
(remaining) end-points of the two original clusters. The number
of original sites that belong to a cluster defines its
`magnetic moment' and it then makes sense to talk of a
magnetization density $m$ for the system.  
Pictorially, a reference dimer cover specifies some 
connectivity rules for the RG-formed dimers, with natural 
notions of clusters and percolation with respect to such 
connectivity rules, while the RG rules
prescribe the dynamics (as a function of increasing $\Gamma$)
obeyed by the clusters.  
Such connectivity properties can be used to distinguish 
between different RG-generated dimer covers, and, thus, 
to characterize the different phases.\cite{stringorder} 
Within the strong-disorder RG, this is the distinction 
that captures the different `topological' character of 
the two Griffiths phases.

\begin{figure}
\epsfxsize=\columnwidth
\centerline{\epsffile{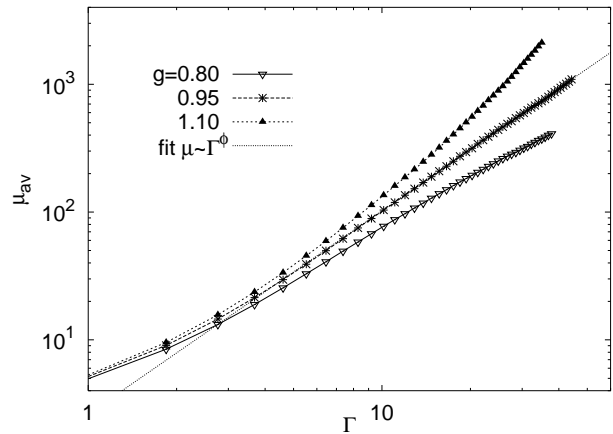}}
\vspace{0.15in}
\caption{Average moment of the remaining clusters with respect 
to the vertical reference cover.  Also shown is the fit at
the critical point for the exponent $\phi$.  From
this fit we obtain $\phi \!\approx\! 1.59$, which can
be compared with the corresponding exponent for the quantum
Ising model, $\phi_{\mathrm RTFIM}=(1+\sqrt{5})/2 \approx 1.62$.
}
\label{RGKM2mu}
\end{figure}

Going back to our numerical RG studies, we find that,
with respect to the vertical reference cover, in the
phase which obtains for strong rung couplings (small values
of $g$) there are only small clusters and no connectedness across 
the whole system, while for weak rung couplings (large values
of $g$) there is an infinite percolating cluster that forms
in the limit of large $\Gamma$.  This development
of topological order for $g>g_c$ is characterized by
an exponent $\beta$ defined by the scaling of the
average magetization density:
$m(\Gamma \rightarrow \infty) \sim (g-g_c)^\beta$.
At criticality, the average moment of the surviving clusters scales 
as $\mu \sim \Gamma^\phi$, which 
defines the exponent $\phi$; in complete analogy to the 
RTFIM,\cite{dsf} the exponent $\beta$ for the topological
order parameter (`magnetization') can be obtained 
from $\phi$ via the scaling relation $\beta=2-\phi$.
Figure~\ref{RGKM2mu} shows our numerical result
for the $\Gamma$ dependence of the average 
moment of the surviving clusters at the critical point.
The numerical value obtained for the exponent
$\phi$ is very close to that of the $1d$ RTFIM. The value
$\beta \approx 0.41$ we infer using the scaling relation is then
very close to the corresponding exact result for the single dimerized chain
(which has a $\beta$ exactly equal\cite{hybg} to the magnetization exponent
$(3-\sqrt{5})/2$
of the $1d$ RTFIM~\cite{dsf}).
We also note that, similarly to the quantum Ising model,
the critical point is the point of balance between 
the cluster fields and cluster interactions; this is 
shown in the inset of Fig.~\ref{RGKM2isolani}, and provides
alternative means for identifying the critical point.
This completes our RG description of the spinless case.

\narrowtext

Essentially the same results are obtained for the corresponding 
spinful problem with T-symmetry---this is consistent with our 
general argument in Section~\ref{strongrgrules}.
We also repeated this analysis for the specific realization of 
the spinful problem without T-symmetry corresponding to 
Fig.~\ref{phased2}{\tt A}.  All the results obtained for this 
case in the vicinity of the transition at fixed finite 
dimerization (as a function of increasing $R$) are essentially 
identical to the results shown above for the spinless case.
[Entirely analogous results are also obtained in all cases for the 
dimerization driven transitions of the bipartite ladder problems.]
Thus, our general picture for the low-energy physics appears to be
validated by the RG results so long as we are not in the vicinity
of any special multicritical points.

Finally, a brief comment on the RG results in the vicinity of 
the putative multicritical point, Fig.~\ref{phased2}{\tt C}. 
As we scan R at $\delta=0$, strong Griffiths effects again show up 
clearly over a wide region in the vicinity of the putative 
multi-critical point.  Moreover, the phases at large and small R 
both look `paramagnetic' in terms of clusters defined with
respect to the vertical reference cover (and also many other 
reference covers).  This is consistent with our arguments 
in the previous section.  However, our RG 
results are also inconclusive when it comes to pinning down 
the structure of the phase diagram near this apparent 
multicritical point---again the analysis is plagued by
near-critical behavior over a wide region.  The corresponding
long crossovers do not allow us to make any definitive statements
regarding either the presence or the universality class of such
multicritical points.  This remains the principal unresolved 
question in our entire analysis of the one dimensional examples.

\section{How are the low-energy states generated?}
\begin{figure}[!t]
\epsfxsize=\columnwidth
\centerline{\epsffile{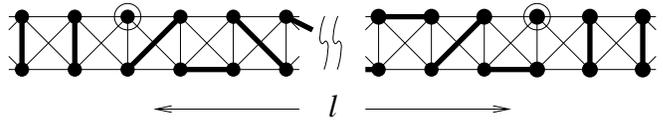}}
\vspace{0.15in}
\caption{Griffiths generation of low-energy states.
}
\label{Griff}
\end{figure}

We now consider precisely how the states in the singular 
low-energy tail of the density of states are generated, and
identify the corresponding Griffiths regions in the ImRH 
language.  As is already implicit in the above discussion,
such a pair of low-energy states is formed if there are two
`isolated' sites separated by a region in which all the other
sites are locked into short-ranged (dimer) pairs. 
Griffiths effects are a consequence of such isolated sites 
being generated sufficiently often.  An example of a pair of 
such sites is shown in Fig.~\ref{Griff}.  The isolated sites 
are very weakly coupled to each other---the coupling is of order 
$\epsilon \sim e^{-c\, l}$, where $l$ is the length of 
the region.  In a disordered system, there is always 
a probability of order $p^{\,l}$, with some $p<1$, of finding 
such a region---this gives a power-law contribution 
$\sim \epsilon^{\,|\!\ln p|/c}$ to the integrated density of 
states from such regions.  Thus, we {\it always} expect a
variable power-law (Griffiths) density of states in such
random hopping problems in which {\it no on-site energies}
(i.e., diagonal terms) are allowed; we conclude that Griffiths 
phases are {\it generic}.  The specific example shown in the figure
is expected to be relevant to Griffiths effects in
the phase in which the rung couplings dominate on average.  
The regions to the left and to the right of the pair 
are intended to be a caricature of the {\it typical} 
regions in this regime, while the region 
in the middle is comparatively {\it rare}.  Decreasing the 
strength of the rung couplings would increase the probability 
of having such regions, corresponding to the observed increase 
in the dynamical exponent as one approaches the transition to 
the other phase.  The critical point is then characterized by 
a proliferation of such Griffiths regions on all energy scales.

The picture that emerges is thus very similar to that in 
our toy-model of Sec.~\ref{toy} in which the low-energy 
states are closely associated with `domain walls' between
the two different gapped phases of the pure system.

\narrowtext

It is useful, at this point, to recast some of this in terms 
of the original superconductor Hamiltonian $\hat{H}_{SC}$.
This will give us a somewhat different perspective on the origin
of these low-energy states.
For simplicity, we restrict the discussion below to the spinless 
case.  We first examine the basic RG rule Eq.~(\ref{RGrule}).  
Consider a two-site problem
\begin{eqnarray}
\hat H[\alpha,\beta] = 
\epsilon_\alpha c_\alpha^\dagger c_\alpha +
\epsilon_\beta c_\beta^\dagger c_\beta +
(t c_\alpha^\dagger c_\beta +
\Delta c_\alpha^\dagger c_\beta^\dagger + {\mathrm h.c.}).
\nonumber
\end{eqnarray}
This can be diagonalized by an appropriate Bogoliubov 
transformation to give~\cite{levitov}
\begin{eqnarray}
& \hat H[\alpha,\beta]  =  \epsilon_+ \gamma_+^\dagger \gamma_+ +
\epsilon_- \gamma_-^\dagger \gamma_- + {\mathrm const},
\nonumber \\
& \epsilon_{\pm} = \frac{1}{2} \left|
\sqrt{(\epsilon_\alpha + \epsilon_\beta)^2 + 4|\Delta|^2} \pm 
\sqrt{(\epsilon_\alpha - \epsilon_\beta)^2 + 4|t|^2} 
\right| . \nonumber
\end{eqnarray}
From this solution, it is clear that states at both $\alpha$ and 
$\beta$ simultaneously contribute to either eigenstate only if
either $|t| \agt |\epsilon_\alpha - \epsilon_\beta|$ or
$|\Delta| \agt |\epsilon_\alpha + \epsilon_\beta|$.
When this happens, we can no longer think of the two sites 
in isolation and need to account for the {\it resonance}
between them.  Now, if 
$|t|^2-|\Delta|^2 \sim \epsilon_\alpha \epsilon_\beta$,$\,$
$\epsilon_-$ for such a resonance can be very small.
In order to obtain a good low-energy description in such 
a situation, we would eliminate the high energy `$+$' state 
and keep only the `$-$' state by introducing a single 
{\it effective} site with site energy $\epsilon_-$.  In our 
ImRH RG language, this corresponds to a situation in which 
a single bond connecting sites on two different rungs is 
decimated, leaving behind one site on each rung; these two 
remaining sites are coupled by a weak bond precisely equal 
in magnitude to $\epsilon_-$.  On the other hand, if 
there is no mixing of the states at $\alpha$ and 
$\beta$, i.e., when one of the local potentials, 
say $\epsilon_\alpha$, dominates, we would eliminate the 
state $\alpha$ completely; in the ImRH RG this corresponds
to decimating the corresponding rung.
Thus, our RG procedure either eliminates a full fermion state
because it is frozen out by a strong local potential, or 
eliminates `half of a fermion' from each of the two sites in 
resonance and recombines the remaining halves to create a new 
effective fermion with site energy equal to the new coupling 
introduced (our RG is thus really defined on the corresponding 
Majorana fermion states). 

Now, consider for concreteness the Griffiths phase in which 
the on-site potentials dominate.  At low enough energies, 
a description in terms of isolated effective sites (with 
negligible mixing between them) with some renormalized 
distribution of effective site energies is clearly appropriate.  
However, to arrive at such a description, one has to first 
account for all the resonances at higher energy scales 
that arise from any anomalous regions in which hopping and pairing 
amplitudes are large compared to local potentials.  As a specific 
example of such anomalies, consider the central region of 
Fig.~\ref{Griff}.
In the original superconductor language, this region 
corresponds to both $\Delta$ and $t$ being relatively large 
and comparable in magnitude, in addition to having a somewhat 
definite relationship between their phases throughout this 
region.  Eliminating all the resonances between the sites in this 
region finally gives a low-energy description in terms of 
a single effective site with an exponentially small energy. 
This effective site in the original superconductor language 
clearly corresponds, in the ImRH language, to the pair of 
isolated sites shown in Fig.~\ref{Griff}.
The ImRH RG thus provides the natural language for
capturing the important low-energy physics.  Crudely 
speaking, the `effective site-energies' in the original 
superconductor language correspond to the `cluster field
couplings' with respect to the natural vertical reference 
dimer cover in the ImRH RG language.  Note also that an 
important ingredient of the physical picture that emerges is 
the spatial character of the quasiparticle states (in particular,
note that wavefunction of a low-energy quasiparticle
in the phase with zero-energy end-states is split into two spatially 
separated pieces).  Such spatial information 
is also kept most naturally in the ImRH RG; in particular, 
the development of the `topological order' is seen most 
naturally in this language.

Finally, we can now go back and ask what is the precise role 
played by the various symmetry restrictions.  As discussed above,
Griffiths effects in the ImRH RG language are associated with 
situations in which we repeatedly eliminate only a subset of 
the couplings connecting two rungs (for instance, only 
{\it one} of the couplings in the spinless example above), 
thus splitting the original fermion states.  
Now, if some symmetry restrictions require that 
some of the original couplings between rungs equal one another, 
then we would be forced to decimate several of them 
simultaneously---for example, we could be forced to 
eliminate only complete fermion states.  Within the RG 
approach, this could then rule out the possibility 
of Griffiths effects.  It is easy to see that the restrictions 
imposed by T-invariance are not enough for this to happen. 
On the other hand, {\it in the SR-invariant case}, a simple 
analysis of the symmetry constraints for the corresponding 
ImRH problem shows that this is precisely what happens.\cite{normal} 
In the strong disorder limit, this, then, is the true 
significance of the absence or presence of SR-invariance 
for quasiparticles of a superconductor.

\section{Discussion}
In this article, we have established the existence of strong 
Griffiths effects in one-dimensional superconducting wires in 
which the quasiparticle spin is not a good quantum number.
We associated these singularities with quasiparticle states
that live on `domain walls' between adjacent large regions
in two different phases, one phase that supports zero modes
localized near the ends of a finite system, and another
that does not support such modes.

An obvious question now arises:  Do such effects exist in 
two or more dimensions?  Thinking in terms of the ImRH RG, it does 
seem that such effects could exist in cases without SR-invariance,
particularly in the vicinity of the thermal-metal to 
thermal-insulator transition, or the transition between the two 
topologically distinct insulating phases in cases without 
T-symmetry.  However, it is of course not {\em a priori} 
clear if `isolated sites' would be produced sufficiently
often in the RG for this to happen.  Moreover, the extent
to which such isolated sites can be associated, as in one 
dimension, with domain walls between topologically distinct 
regions is also not clear.
Another obvious question is whether any of the phase transitions 
mentioned above are controlled by infinite-disorder fixed points.
Although we have not studied all these questions carefully, 
it is clear that our strong-disorder RG approach, implemented 
numerically, provides the natural tool for such investigations
in more than one dimensions.
[So far, we have extensively studied only a related $2d$ 
bipartite random hopping problem\cite{gade} using a similar 
RG approach\cite{Rgcomment} and found that such Griffiths 
effects do occur in this case:  When we introduce enforced 
dimerization into this random hopping problem, we observe a 
transition between different Griffiths phases and tentatively 
conclude that the critical point is controlled by an 
infinite-randomness fixed point.  Details of this study will 
be reported elsewhere\cite{usgade} and could be relevant for 
certain models for spinless T-invariant $2d$ superconductors 
in the presence of disorder.]

\section{Acknowledgements}
We would like to thank F.~D.~M.~Haldane, B.~I.~Halperin, 
M.~Hastings, I.~Gruzberg, R.~Moessner, and A.~Vishwanath 
for valuable discussions, and C. Mudry and T. Senthil for 
useful comments on the manuscript.  One of us (KD) was supported by 
NSF grant DMR-9809483 while at Princeton and is currently 
supported by NSF grant DMR-9981283 at Harvard.  The others 
acknowledge support of NSF grant DMR-9802468.


\begin{references}
\bibitem{ilyafoot} After this paper was completed, we learned
that the generic localized behavior predicted by us can also 
be studied using the supersymmetry method for the special case 
of a 1d single-channel spinless superconductor without 
T-invariance (I. Gruzberg {\it et. al.}, private communication).
\bibitem{foottop} This distinction between the two phases
of the pure system can also be stated in topological
terms as in Ref.~{\protect{\onlinecite{rg}}}.
\bibitem{Rgcomment}  Note that this is a natural generalization
of the RG transformation $\tilde t_{iA,jB} = 
t_{iA,jB}-t_{iA,2B} (t_{2B,1A}^*)^{-1} t_{1A,jB}$ introduced in
previous work~{\protect{\cite{mdh}}} to treat a bipartite random 
hopping problem (in which sites of the `A' sublattice are only 
coupled to those of the `B' sublattice).  In this case too,
no diagonal terms are generated and the problem remains 
bipartite.  However, for a general hopping problem (not pure 
imaginary) on a non-bipartite lattice the transformation 
Eq.~({\protect{\ref{RGrule}}}) generates diagonal terms 
(corresponding to random site energies), and the system is 
in the usual universality class of Anderson localization.
\bibitem{normal} The same is true if we write the usual 
Anderson localization problem in the ImRH language---this 
is consistent with the expectation that disorder does not lead 
to any singular behavior of the density of states at the Fermi
energy in a normal, non-interacting electron system.
\end{references}
\end{document}